\documentstyle[preprint,aps]{revtex}
\begin{document}
\preprint{KAIST-TH 9/96 , SNUTP 96/047}

\title{\bf Axion Cosmology with a Stronger QCD in the Early Universe}

\author{Kiwoon Choi}

\address{Physics Department, Korea Advanced Institute of Science
and Technology, Taejon 305-701, Korea}

\author{Hang Bae Kim}

\address{Departamento de F{\'\i}sica Te\'orica C-XI,
Universidad Aut\'onoma de Madrid,\\ Cantoblanco, 28049 Madrid, Spain}

\author{Jihn E. Kim}

\address{Department of Physics and Center for Theoretical Physics,
Seoul National University, Seoul 151-742, Korea}

\maketitle

\begin{abstract}
We examine in the context of supersymmetric
models  whether the usual cosmological upper bound on the axion
decay constant can be relaxed by assuming a period of stronger
QCD in the early universe.
By evaluating  the axion potential in the early
universe
and also taking into account the dilaton potential energy,
it is argued that a stronger QCD 
is not useful for raising up
the bound.

\end{abstract}
\pacs{}
\newpage
\def\be{\begin{equation}}
\def\ee{\end{equation}}

\section{Introduction}

One of the most attractive solutions to the strong CP problem is
to introduce a spontaneously broken Peccei-Quinn symmetry \cite{pq}.
This solution originally predicted a pseudo-Goldstone boson, the so-called
Peccei-Quinn-Weinberg-Wilczek axion \cite{wein}, which is ruled 
out phenomenologically.  The subsequently devised 
invisible axion \cite{invisible}
is also constrained to a narrow band on the axion decay constant $f_a$.
Some astrophysical arguments,  e.g. those associated
with the axion emission from  helium burning red giants or the 
supernova SN1987A \cite{kim,raffelt}, imply that $f_a$ is far above the 
weak scale.  A natural possibility is then that $f_a$ is around 
the grand unification scale or the Planck scale.
As is well known, however, the coherent axion
oscillation in the early universe gives rise to relic axions whose  
mass density $\Omega_a$ (in the unit of the critical
energy density)  at present is given by \cite{preskill}
\be
\Omega_a=(\delta a/f_a)^2(f_a/4\times 10^{12} \, {\rm GeV})^{1.18},
\ee
where $\delta a$ is the 
axion misalignment at the epoch of the QCD phase transition.
Clearly it is most natural to assume that the axion misalignment is
of order  $f_a$. Requring  the axion mass density not significantly
exceed the critical density, this leads to  the  upper bound
$f_a\leq 4\times 10^{12}$ GeV which is far
below the grand unification scale or the Planck scale.

One way to raise up the cosmological upper bound on $f_a$ is to have an entropy
production after the QCD phase transition \cite{turner}. 
However it has been argued that, due to  constraints from
the nucleosynthesis, one cannot
reach above few times $10^{15}$ GeV by this late time entropy
production \cite{seckel}.
Recently another interesting way to raise up the axion bound 
was suggested by Dvali \cite{dvali}. 
The idea is to suppress the axion misalignment
$\delta a$  by assuming a period of  stronger QCD in the early universe
during which the QCD coupling constant takes a value  larger than
the present one.
Clearly this would be possible only when 
the QCD coupling constant is determined by a dynamical degree of freedom in
the theory, e.g.  a dilaton field $\phi$ whose vacuum expectation value
(VEV)
gives   
$1/g^2=\langle \phi\rangle$.
If the dilaton takes a value smaller than the present one,
the QCD at that time would be stronger, leading to
the corresponding scale $\Lambda_{QCD}$ {\it far bigger}
than the present value of order $0.2$ GeV.
Then there is a possibility that a large
effective axion mass is induced in the early universe,
driving  the axion field toward the minimum of the effective potential.
If the minimum of the early axion potential 
coincides with the minimum of the present axion potential,
$\delta a$ would be dynamically suppressed, allowing the upper bound of
$f_a$ larger than $4\times 10^{12}$ GeV. Note  that 
Eq. (1) indicates that the upper bound of $f_a$
can be around $10^{16} \sim 10^{18}$ GeV if $\delta a/f_a$
is relaxed down to $10^{-2}\sim 10^{-3}$ in the early universe.  

In order for 
the axion misalignment $\delta a$ to be relaxed by
the above mechanism,
one needs
\be
(i) \quad m_a \geq H, \, \, \quad 
(ii) \quad |\langle a/f_a \rangle -\theta_{\rm eff}| \leq 10^{-3}\sim 10^{-2},
\ee
where  $m_a$ and $\langle a\rangle$ denote the axion mass and
the vacuum expectation value (VEV) determined by the axion potential  
in the early universe, and  $\theta_{\rm eff}$ is the low energy 
QCD vacuum angle {\it at the present}.  Here  the first condition 
is required for the onset of the axion field oscillation 
toward the minimum, and the second is for 
the location of this minimum to coincide
with the present one up to a small misalignment angle
of order $10^{-3}\sim 10^{-2}$. Note that the present axion VEV is given by 
the low energy vacuum angle $\theta_{\rm eff}$
which includes the contributions from  a variety of
CP-odd parameters in the underlying theory.   For instance, 
if the underlying theory is the minimal supersymmetric standard model,
\be
\theta_{\rm eff}=
\theta_{QCD}+{\rm arg}({\rm det}(\lambda_u\lambda_d))+
3{\rm arg}(m_{1/2})-3{\rm arg}(\mu B),
\ee
where $\theta_{QCD}$ denotes the {\it bare} QCD vacuum angle,
$\lambda_u$ and $\lambda_d$ are the Yukawa coupling matrices, 
and $m_{1/2}$, $\mu$ and $B$ denote the gluino mass, the Higgs $\mu$
parameter, and the soft $B$ parameter, respectively.
We stress that  the second condition in Eq. (2) 
for the early minimum to coincide with
the present one  should be satisfied  {\it without} fine tuning of
some field values which would be {\it chaotic} in the early universe.
Such fine tuning
is certainly nothing better than the simpler  fine tuning
of the chaotic
axion
field  $a/f_a$    to $\theta_{\rm eff}$ at the epoch of
QCD phase transition.

In this paper, we wish to examine whether
the axion misalgnment can be relaxed  by a stronger
QCD in the early universe
in  supersymmetric models.
To start with, we assume that
the dilaton dynamics in the model allows a period
in the early universe during which QCD becomes stronger.
The reason for limiting the analysis to supersymmetric models
is clear:  the most natural playground of the dilaton field  
is supersymmetric model.

The organization of this paper is as follows.
In Section 2, we provide a discussion of the axion potential
in the early universe
by distinguishing the two cases, one in which $\Lambda_{QCD}$ is smaller
than the size of soft supersymmetry breaking parameter
$m_{\rm soft}$ and the other opposite case that
the QCD becomes so strong that  $\Lambda_{QCD}\gg m_{\rm soft}$.
We discuss the second case in somewhat detail since the axion potential
in this case has not been considered in the previous
literature.  As we will see, the axion potential in the second case 
dramatically differs from  the well known axion
potential in the first case. Based on this analysis,  in Section 3
we examine whether the two conditions given
in Eq. (2) can be simultaneously met.
Taking into account the dilaton potential energy,
we will argue that it is very {\it unlikely} to have a 
cosmological scenario in which both of 
these two conditions are  satisfied.
We thus conclude in Section 4 that a stronger QCD in the early
universe is {\it not}  useful for raising the cosmological
upper bound of the axion decay constant $f_a$.

\section{Axion Potential in the period of stronger QCD}

In this section, we present a somewhat detailed
discussion of the axion potential in the early universe.
In the early universe, field variables would take chaotic values 
if their effective masses
are smaller than the expansion rate $H$.
In order to satisfy Eq. (2) without fine tuning,
one needs to keep the axion VEV {\it not} shifted by such  chaotic fields.
If not,  the axion VEV in the early universe
would become chaotic and thus
not coincide with   the present axion VEV in general.
We will thus assume that field variables which  apparently
affect the axion VEV are {\it not} chaotic, {\it but}
quickly settle  into the minimum
of their effective potential in the early universe.
Of course, one still needs to confirm that these  fields
settled at the minimum leads to an early axion VEV almost the same as 
the present one as given in Eq. (2).  

To compute the axion potential in the early universe,
it is convenient to distinguish the following
two cases: 
$$
(i) \quad \Lambda_{QCD} < m_{\rm soft},  
\quad \quad (ii) \quad \Lambda_{QCD}\gg m_{\rm soft}.
$$

Let us first consider the case that  $\Lambda_{QCD} < m_{\rm soft}$.
In this case, if there is no light quark whose current mass 
is less than $\Lambda_{QCD}$,
the axion potential is given by
\be
V_a\simeq \Lambda_{QCD}^4 \cos (a/f_a-{\theta_{\rm in}}), 
\ee
where  ${\theta_{\rm in}}$ denotes the QCD vacuum angle
{\it in the early universe}.
The presence of light quarks suppresses the axion
potential as
\be
V_a\simeq
m_q\Lambda_{QCD}^3 \cos (a/f_a-{\theta_{\rm in}}),
\ee
where $m_q$ $(< \Lambda_{QCD})$ is the  current mass of
the  {\it lightest} quark.
 
Of course, it depends upon the Higgs VEVs
in the early universe whether there exists a light quark
with $m_q<\Lambda_{QCD}$.
In fact, the Higgs VEVs in the early universe can dramatically
differ from the present ones.
To be explicit, let us consider the minimal supersymmetric
standard model.
In this model, the Higgs potential for
the neutral components takes the form
\be
m_1^2|H_u|^2+m_2^2|H_d|^2+(B\mu H_uH_d+{\rm h.c.})
+\frac{1}{8}(g_2^2+g_1^2)(|H_u|^2-|H_d|^2)^2,
\ee
where $H_u$ and $H_d$ denote the two Higgs doublets in the model.
In generic supersymmetric models,
soft parameters are  determined by the VEVs of 
fields triggering spontaneous SUSY breaking.
As a result, soft parameters in the early universe
may take values which differ from the present ones.
In particular, both the sign and magnitude of  $m_1^2$ and $m_2^2$ 
in the early unverse can differ from the present ones
due to an additional SUSY breaking
by nonzero energy density.
This change of $m_1^2$ and $m_2^2$
does  not alter ${\rm arg}(H_uH_d)$
and thus {\it not} lead to a  change of 
the early axion VEV.
However it can dramatically change $|H_u|$ and $|H_d|$.
For instance, if both $m_1^2$ and $m_2^2$ are positive and large
enough, the Higgs VEVs do vanish.
Then all quarks become massless
and the axion potential vanishes up to
a tiny contribution from small instantons.
In the opposite case that both $m_1^2$ and $m_2^2$ are negative,
the Higgs VEVs slide down toward the flat direction $|H_u|=|H_d|\gg
m_{\rm soft}$.
In this case, we  would not have any light quark and
the axion potential is given by (4).
Finally for $m_1^2$ and $m_2^2$ leading to the Higgs VEVs
similar to the present ones,
the axion potential
is given by (5) with  the up quark mass
$m_q\simeq 10$ MeV.

In the above, we have ignored a possible finite temperature effect
on the axion potential.
At high temperature $T > \Lambda_{QCD}$,
the axion potentials Eq. (4) and Eq.(5) are {\it not}
valid anymore.  However, compared to the low temperature result,
the high temperature  axion potential is  suppressed
by  more powers of the light quark masses and also by 
$\exp (-8\pi^2/g^2(T))$ \cite{gross}.  Since what we wish to obtain 
is a bigger axion potential for a given value of $\Lambda_{QCD}$,
we will not consider such a high temperature case  anymore.

In the above discussion, the Higgs fields were assumed to
stay at the minimum of the potential.
It has been argued that in generic supergravity models
a nonzero energy density $\rho_E\sim H^2M_{\rm Pl}^2$ in
the early universe  affects soft parameters as  $\delta m_{\rm soft}=c H$
where $H$ is the Hubble expansion rate and  $c$ is a dimensionless constant
\cite{dvali1}. The constant $c$ may be small due to small couplings
or to loop suppression. If $c$ is small enough, it is possible that
$m_{\rm soft}\ll H$.  Such a large $H$ may  result in
chaotic Higgs values in the early universe.
However, in supersymmetric models with two Higgs doublets $H_u$ and $H_d$,
chaotic Higgs fields lead to a chaotic axion VEV
which differs from the present axion VEV in general.
Furthermore, for  $H\geq m_{\rm soft}$,
we have $m_a\ll H$ (Note that we have been discussing    the case
$\Lambda_{QCD}< m_{\rm soft}$.) and thus the first condition
of Eq. (2) can  {\it not} be satisfied also.  We thus assume that
$H\leq m_{\rm soft}$
and thus the Higgs  fields quickly settle into their VEVs.

Let us now consider the opposite limit
that  the QCD becomes so strong
that $\Lambda_{QCD}\gg m_{\rm soft}$.
Then the axion potential is induced by nonperturbative 
supersymmetric QCD effects.
Recent progress in understanding the dynamics of
supersymmetric gauge theories \cite{seiberg}
allows us to estimate 
the axion potential even in this case.

As is well known, the axion potential is severely affected
by  gauge invariant  condensates which  break chiral symmetries
in the theory.
These condensates can be used to tie together fermion zero modes 
of the instanton amplitudes 
generating the axion potential.
Phase degrees of freedom of such condensates then  mix
with the axion field as the $\eta^{\prime}$ mixes with
the axion in normal QCD \cite{kim}.
We thus  study first what kind of chiral symmetry
breaking condensates are formed 
in the supersymmetric QCD with the superpotential
\be
W=\lambda_u H_uQu^c+\lambda_d H_dQd^c,
\ee
where $Q$, $u^c$, and $d^c$  denote 
$SU(2)_L$ doublet quarks and singlet antiquarks, respectively.
Here, the generation indices and gauge group indices are 
omitted for simplicity.

It has been argued that, for a vanishing superpotential,
the quantum moduli space 
of degenerate vacua   for
supersymmetric $SU(3)_c$ gauge theory with $N_f=6$
quark flavors is the same as the classical one \cite{seiberg}. 
Such a vacuum degeneracy is lifted in fact by the Yukawa terms
in the superpotential and also by soft breaking terms \cite{peskin}.
As was observed recently, the infrared behavior of gauge
invariant operators in
supersymmetric $SU(N_c)$ gauge theory with $N_f$ quark
flavors 
can be studied  by the dual 
$SU(N_f-N_c)$ theory with
$N_f$ dual quarks \cite{seiberg}.
The superpotential of the dual model contains 
\be
W_D=\Lambda_{QCD}(\lambda_u H_uT_u+\lambda_d H_d T_d)+
T_uQ_Du_D^c+T_dQ_Dd_D^c,
\ee
where $T_u=Qu^c/\Lambda_{QCD}$ and $T_d=Qd^c/\Lambda_{QCD}$ denote
the meson superfields with dimension one at the ultraviolet, 
while $Q_D$, $u_D^c$,
and $d_D^c$ are the dual quark and anti-quark superfields. 
Once the supersymmetry of the original model is broken by
soft parameters much less than $\Lambda_{QCD}$,
the dual model will contain soft terms \cite{peskin}
\be
{\cal L}_{\rm soft}^{(D)}=AW_D+\sum m_I^2|\phi_I|^2,
\ee
where  $\phi_I$ denotes generic scalar fields
in the dual model and the soft parameters
$A$ and $m_I$ are again   much less than $\Lambda_{QCD}$.

A nice feature of the dual theory is that it becomes weaker
as  the original QCD becomes stronger, allowing a classical
approximation to be valid.  The classical  effective
potential of the dual model can be readily computed
from the superpotential (8) and the soft terms (9).
It is then easy to see that, for  $\Lambda_{QCD}$  large
enough compared to $m_{\rm soft}$,  the Higgs
fields $H_{u,d}$ and the linear combinations
$\lambda_uT_u$ and $\lambda_dT_d$
of mesons have  {\it positive} mass squared
of  order  $\Lambda_{QCD}^2$, and thus have {\it vanishing} VEV's.
For the original squarks having {\it positive} soft mass squared,
it is expected that both the mesons and the dual squarks
also have positive soft mass squared.
Note that the mesons are  the bound states
of the original squarks and anti-squarks,
while the dual squarks can be obtained 
by dissociating the scalar baryons (containing  $N_c$ 
original squarks) into $(N_f-N_c)$ pieces.
With  positive soft mass squared, the VEVs of the dual squarks and
the entire mesons {\it vanish} also.

The above results on the VEVs 
can be summarized as
$$
\langle \tilde{q}\tilde{q}^c\rangle\propto \langle T\rangle=0,
$$
$$
\langle q q^c\rangle\propto\langle F_T\rangle =0,
$$
\be
\langle \lambda\lambda\rangle 
\propto \langle T\rangle^{N_f/(N_f-N_c)}=0,
\ee
where $T$ denotes the scalar components of the meson superfields
$T_{u,d}$ with the
auxiliary components $F_T$.
Thus all of the squark condensates
$\langle \tilde{q}\tilde{q}^c\rangle$, the quark condensates $\langle
qq^c\rangle$,
and the gluino condensate $\langle\lambda\lambda\rangle$ do vanish.
Note that the vanishing of the gluino condensate is essentially
due to the vanishing Higgs VEV leading to the massless quarks.
Although the above chiral invariant vacuum configurations
were derived  based on the classical potential at zero temperature,
it is rather easy to see neither quantum corrections (due to the weak
dual gauge interactions) nor 
finite temperature effects does  shift these 
chiral invariant configurations.

So far, we have argued  that
the mesons and the dual squarks have vanishing VEVs
for $\Lambda_{QCD}\gg m_{\rm soft}$.
Again if the expansion rate $H$ is large enough so
that $H\gg m_{\rm soft}$,
the mesons and dual squarks would have chaotic
values although their effective potential is minimal
at zero values. Note that the dual squarks
and the  mesons (except for $\lambda_uT_u$ and $\lambda_dT_d$)
have masses of order $m_{\rm soft}$.
Chaotic values of  the mesons and the dual squarks  correspond 
to  chaotic condensates of the quarks, squarks, and gluino.
Clearly the axion VEV determined by such chaotic condensates  
will be {\it chaotic} also, and thus does {\it not} coincide in general
with the present axion VEV.  Thus to avoid a chaotic axion VEV,
we assume that $m_{\rm soft}\geq H$ and 
the mesons and the dual squarks quickly settle into
the minimum of their effective potential.

In the case that there is {\it no} condensate,
the axion potential is given by the QCD instanton amplitudes
whose fermion zero modes are tied together by the interactions
which break the chiral symmetries explicitly \cite{flynn}.
To proceed, let us consider  
the minimal supersymmetric standard model having the superpotential
\be
W=\lambda_uH_uQu^c+\lambda_dH_dQd^c+\mu
H_uH_d,
\ee
and the soft terms
\be
{\cal L}_{\rm soft}=
\frac{1}{2}m_{1/2}\lambda\lambda+A(\lambda_u H_u\tilde{Q}\tilde{u}^c
+\lambda_d H_d\tilde{Q}\tilde{d}^c)+B\mu H_uH_d +\frac{1}{2}\sum m_i^2
|\phi_i|^2+{\rm h.c.},
\ee
where $\phi_i$ denotes generic scalar fields in the model.
In order to compute the instanton-induced
axion potential,  we first note that  the model is invariant
under  
\be
G_{MSSM}= SU(3)_Q\times SU(3)_{u^c}\times SU(3)_{d^c}
\times U(1)_A\times U(1)_X\times U(1)_R,
\ee
where the fields and parameters transform as
$Q=(3, 1,1)$, $u^c=(1,\bar{3}, 1)$, $d^c=(1,1,\bar{3})$,
$\lambda_u=(\bar{3},3,1)$, and $\lambda_d=(\bar{3},1, 3)$
under $SU(3)_Q\times SU(3)_{u^c}\times SU(3)_{d^c}$,
and the quantum numbers of $U(1)_A\times U(1)_X\times U(1)_R$
are given in Table 1.  As is well known,
the soft supersymmetry breaking can be described within superspace
formalism by introducing  spurion superfields:
$$\eta=\{1+m_i^2\theta^2\bar{\theta}^2\},$$ 
$$Y=(1+16\pi^2m_{1/2}\theta^2)\tau,$$
$$Z=\{ Z_{u,d}=(1+ A\theta^2)\lambda_{u,d}, \quad
Z_{\mu}=(1+ B\theta^2)\mu\},$$
where the auxiliary components of the spurion superfields
represent soft breaking, while the scalar components
denote supersymmetric couplings: the complex gauge coupling
$\tau=\frac{8\pi^2}{g^2}+i\theta_{QCD}$, the Yukawa couplings
$\lambda_{u,d}$, and the $\mu$ parameter.
Note that the factor $16\pi^2$ in $Y$ 
is introduced for a  proper normalization of the gluino mass $m_{1/2}$.
Obviously  $\eta$'s are 
real superfields while $Y$ and $Z$'s  are chiral.

After integrating out the gauge and matter superfields, the axion effective
Lagrangian can be read off from the  effective lagrangian of
the spurions by identifying
the dimensionless axion superfield 
$$
A=(s+ia+\tilde{a}\theta+ F_A\theta^2)/f_a 
$$
as the fluctuation of $Y$, i.e. by the identification
\be
Y\rightarrow Y+A.
\ee
The effective Lagrangian of spurions  can be written as
\be
\int d^2\theta d^2\bar{\theta} \, 
K_{\rm eff}(Y, Y^*, Z, Z^*, \eta)
+\int d^2\theta \, W_{\rm eff}(Y,Z)+{\rm h.c.},
\ee
where the effective Kahler potential $K_{\rm eff}$ is a real function
of spurion superfields (and also of their supercovariant derivatives)
while the effective superpotential $W_{\rm eff}$ is a holomorphic
function of chiral spurions.

Since the matter and gauge fields
are integrated over a unique ground state preserving chiral symmetries,
the above effective
Lagrangian does {\it not} have any branch cut associated with
the multiplicity of ground states.
Note that in the case with nonvanishing
gaugino condensate $\langle \lambda\lambda\rangle\sim
e^{-i\theta_{QCD}/N}$, 
the corresponding effective Lagrangian contains terms
depending upon $e^{-Y/N}\sim e^{-i\theta_{QCD}/N}$ and thus has $N$ branches
associated with the $N$ degenerate vacua which are
related to each other by the $2\pi$ shift of $\theta_{QCD}$ \cite{banks}.
In our case with a unique chiral invariant vaccum,
the effective Lagrangian 
is a single valued function of $e^{-i\theta_{QCD}}$,
and thus is manifestly periodic under the $2\pi$ shift of $\theta_{QCD}$.

Instantons would induce a term in
 $W_{\rm eff}$ as
\be
e^{-nY} \omega (Z),
\ee
where $n$ is a positive integer corresponding to the instanton winding number.
It is, however, easy to see that the selection rules
of $G_{MSSM}$ does not allow any
holomorphic $\omega$ which is finite at $\mu\rightarrow 0$.
This implies that the  axion potential does {\it not} appear
through  $W_{\rm eff}$, {\it but} through $K_{\rm eff}$.
Again the selection rules of $G_{MSSM}$ implies that 
at the leading order the instanton-induced Kahler potential 
takes the form (for $n=1$): 
\be
K_{\rm eff}\propto e^{-Y}
{\rm det}(Z_uZ_d) {Z_{\mu}^*}^3 F(\eta)+ {\rm h.c.},
\ee
where  $F$ is an arbitrary function of the real superfields $\eta$.

To obtain the axion potential
from the superspace integration of $K_{\rm eff}$, 
we need at least 
either a single insertion of $D_{\eta}$
or the simultaneous  insertions of $F_{Y,Z}$ and $F^*_{Y,Z}$
where $D$ and $F$ denote the auxiliary components
of spurions.
In other words,  supersymmetry (together with the selection
rules of $G_{MSSM}$)
implies that  the  axion potential is suppressed at least by
the two powers of
$m_{\rm soft}=\{m_i,m_{1/2}, A, B\}$.
Note that here we consider instantons whose
scale, i.e. the inverse of the instanton size $\rho$,
is in the range between $m_{\rm soft}$ (or $\mu$)
and the messenger scale $M_m$ above which soft breaking
is {\it not} operative anymore.
In the popular hidden sector models  in which supersymmetry
breaking is transmitted by supergravity interactions \cite{nilles}, 
the messenger
scale $M_m=M_{\rm Pl}$, while it can be much lower in visible
sector models \cite{nelson}.  
Obviously instantons with $\rho < M_m^{-1}$ (if exist)
do not  contribute to  the axion potential due to 
the restored SUSY.  As we will see, 
instantons which give dominant contributions have
size $\rho\sim \Lambda_{QCD}^{-1}$ and thus belong to the above category
for $\mu\sim m_{\rm soft}\ll \Lambda_{QCD}$.
At any rate, from the K$\ddot{\rm a}$hler potential of Eq. (17), we readily
find  the SUSY suppression factor
\be
[m_{\rm soft}]^2=\{m_i^2,  AB^*,  16\pi^2 m_{1/2} B^*\},
\ee
where
the factor $16\pi^2$ in front of $m_{1/2}$ indicates that
instanton graphs using  $m_{1/2}$ to tie together  gluino zero modes 
contain  one less loop compared to those   using other soft parameters.

With the above observation, one can write the  axion potential
in the case $\Lambda_{QCD}\gg m_{\rm soft}$
as 
\be
V_a = e^{ia/f_a} {\mu^*}^3 {\rm det}(\lambda_u\lambda_d) \, \Omega+
{\rm h.c.}
\ee
where $\Omega$ 
is suppressed by $[m_{\rm soft}]^2$  
and also by some powers of the loop factor $1/16\pi^2$.
One may estimate $\Omega$ using an explicit instanton graph.
For instance, the dimensional
analysis rule of Ref. \cite{flynn}  applied  for the graph of   Fig. 1
yields a rough estimate:  
\be
\Omega\simeq (\frac{1}{16\pi^2})^6\int d \rho \, f(\rho,M)
\, [m_{\rm soft}]^2  \exp 
[-8\pi^2/g^2(\rho)],
\ee
where $[m_{\rm soft}]^2=16\pi^2 m_{1/2}B^*$ and 
$f$ is a {\it dimensionless} function
of the instanton size $\rho$ and the masses  $\{M\}$ of quantum fluctuations
in the graph.
If $\rho M\leq 1$ for all fluctuations, $f$ would be of order unity,
while it is suppressed by
some powers of $1/\rho M$ when  $\rho M\gg 1$ for some of fluctuations.
We already noted that the Higgs masses are of order $\Lambda_{QCD}$.
For $\rho\ll \Lambda_{QCD}^{-1}$, the negative beta function
of supersymmetric $SU(3)$ theory with $N_f=6$ is not negligible,
implying $\exp [-8\pi^2/g^2(\rho)]$ leads to a significant suppression.
For larger instantons with $\rho\gg \Lambda_{QCD}^{-1}$,
although 
 $\exp [-8\pi^2/g^2(\rho)]$ is roughly a constant since the beta function
is small enough to approach to the fixed point \cite{seiberg},
$f$ is suppressed by the large Higgs masses. 
The above arguments imply that the dominant contribution is from 
instantons with size $\rho\sim \Lambda_{QCD}^{-1}$.
We thus have
\be
V_a\simeq e^{ia/f_a} (\frac{1}{16\pi^2})^6
{\mu^*}^3 {\rm det}(\lambda_u\lambda_d)
[m_{\rm soft}]^2\Lambda_{QCD}^{-1}+{\rm h.c.}
\ee
where
$[m_{\rm soft}]^2$ is given in Eq. (18).

In the above, we have computed the axion potential for 
the MSSM with $\Lambda_{QCD}\gg
m_{\rm soft}$.
In fact, the MSSM can be considered as a rather special case 
since it contains a supersymmetric dimensionful parameter $\mu$. 
To see what happens in models without such a parameter,
let us consider 
the next minimal supersymmetric standard model (NSSM)
including an additional gauge singlet
$S$ with the superpotential
\be
W= \lambda_uH_uQu^c+\lambda_dH_dQd^c+\lambda_1SH_uH_d+\lambda_2 S^3.
\ee
As the MSSM, one can use the supersymmetry and also the selection rules of 
$$
G_{NSSM}=SU(3)_Q\times SU(3)_{u^c}\times SU(3)_{d^c}
\times U(1)_A\times U(1)_X\times U(1)_{X^{\prime}}\times U(1)_R,
$$
to constrain the effective Kahler potential $K_{\rm eff}$ leading
to the axion potential.
(For the quantum numbers of $G_{NSSM}$, see Table 2.)
We then find $K_{\rm eff}$ is proportional to
\be
e^{-Y}
{\rm det}(Z_uZ_d) {Z_1^*}^3 D^2Z_2 F(\eta),
\ee
where 
the new spurion superfields $Z_1$ and $Z_2$ are defined as
$$
Z_1=(1+A\theta^2)
\lambda_1, \quad Z_2=(1+A\theta^2)\lambda_2,
$$
and
$D^2=D_{\alpha}D^{\alpha}$ denotes the supercovariant
derivative.
Although $D^2$ is applied to $Z_2$ in the above example, it can be applied
to other spurions also.  For a rough estimate, one may consider
Fig. 2. Again applying the dimensional analysis rule given in
Ref. \cite{flynn}, we find 
\be
V_a\simeq e^{ia/f_a} (\frac{1}{16\pi^2})^8 {\rm det}(\lambda_u\lambda_d)
{\lambda_1^*}^3\lambda_2 A \, [m_{\rm soft}]^2 \Lambda_{QCD}+{\rm h.c.},
\ee
where again $[m_{\rm soft}]^2$ is given in Eq. (18). 

\section{Relaxation of the axion misalignment }

In the previous section, we have estimated the axion potential
by distinguishing the two cases: (i) $\Lambda_{QCD}< m_{\rm soft}$
and (ii) $\Lambda_{QCD}\gg m_{\rm soft}$.
Based on this, in this section we examine whether the axion
misalignment can be relaxed to a small value by a stronger
QCD in the early universe.  To proceed, let us  note that
in generic supergravity models 
some low energy parameters other than $\Lambda_{QCD}$ 
are  determined also by the VEVs of some fields.
In the early universe, such parameters 
may take values which differ from the present ones.
However, in order for the early axion VEV to coincide with
the present VEV, one should require that complex parameters
which affect the axion VEV have almost the same values as 
the present ones.  For the MSSM,  
\be
(\theta_{\rm eff})_{MSSM}=
\theta_{QCD}+{\rm arg}({\rm det}(\lambda_u\lambda_d))+
3{\rm arg}(m_{1/2})-3{\rm arg}(\mu B),
\ee
while for the NSSM we have
\be
(\theta_{\rm eff})_{NSSM}=\theta_{QCD}+{\rm arg}({\rm det}(\lambda_u\lambda_d))
+3{\rm arg}(m_{1/2})-2{\rm arg}(A)-{\rm arg}(\lambda_1^3\lambda_2^*).
\ee
This leads us to assume that in the early universe all 
Yukawa couplings, $\mu$, and complex soft parameters
($A$, $B$, $m_{1/2}$) have the same values as the present ones
in order that  the early axion VEV can
coincide with the present VEV.

Clearly the energy density  in the early universe
contains the dilaton potential energy $V(\phi)$.
At present with $\phi=\phi_0$,
we  have $V(\phi_0)=0$.
However, in the early universe with $\phi\neq \phi_0$
if the dilaton potential is {\it not} flat, which is  believed 
to be the case,
$V(\phi)$ would have a {\it nonzero} positive value.
This means that  raising up
$\Lambda_{QCD}$ to raise up the axion mass $m_a$
raises up also the energy density and thus the expansion
rate $H$.
In supersymmetric models, it is convenient to parameterize 
this dilaton potential energy in the early universe as follows:
\be
V(\phi)=C  m_{\rm soft}^2M_m^2,
\ee
where $C$ is a dimensionless coefficient and
$M_m$ is the messenger scale of SUSY breaking above which
the soft SUSY breaking is not operative anymore and thus a precise
SUSY cancellation takes place.
To be more concrete, here we specify  $m_{\rm soft}$ as the size
of soft SUSY breaking for supermultiplets in the supersymmetric
standard model sector.  

Since it is relevant for our later discussion, 
let us estimate the size of $C$. 
First of all, 
if the dilaton potential is generated directly by the  SUSY breaking dynamics,
$V(\phi)$ is expected to be of order $|F|^2$ where
$F$ denotes the auxiliary components of generic fields in SUSY
breaking sector.
For hidden sector models in which this SUSY breaking
is transmitted by supergravity interactions,
we have $M_m\simeq M_{\rm Pl}$ and $m_{\rm soft}\simeq
|F|/M_{\rm Pl}$ \cite{nilles}, and thus $C$ is of order unity.
For visible sector models in which the SUSY breaking
is transmitted by gauge interactions \cite{nelson}, the messenger scale 
corresponds to the scale of dynamical SUSY breaking and thus
$M_m\simeq |F|^{1/2}$, while  the soft breaking in the
supersymmteric standard model sector is radiatively generated as
$m_{\rm soft}\simeq (\frac{\alpha}{4\pi})^n|F|^{1/2}$ where
$n$ is a model-dependent positive integer.
This implies that $C$ is of order $(4\pi/\alpha)^{2n}$ for
visible sector models when the dilaton potential energy  
directly induced by the SUSY breaking dynamics is parameterized as Eq. (27).
In summary, if the dilaton potential is generated directly
by  the SUSY breaking dynamics,
$C$ would be of order unity or bigger by $(4\pi/\alpha)^{2n}$.

Even in the case that 
the dilaton potential is {\it not} induced directly
by the SUSY breaking dynamics, once SUSY is broken,
it is generated in general by higher order loop effects.
For instance,  loops of colored particles would induce a dilaton-dependent,
i.e. QCD coupling-dependent, contribution to the vacuum energy density.
Such loop effects are quadratic in both the mass splitting in supermultiplets
and the messenger scale $M_m$ which corresponds to the cutoff scale,
and thus can be written as Eq. (27).
(Throughout this paper, the dilaton is minimally defined as  a field 
whose VEV determines the QCD coupling. Of course this dilaton 
can affect other gauge couplings.  Our whole discussion
will be valid also for such general case.)
In the hidden sector models, the mass splitting  is of order $m_{\rm soft}$
which is independent of the QCD coupling at the leading order.
As a result,  a   QCD coupling-dependence 
would  not appear at one loop order,
but does appear at two loops \cite{bagger}. This implies
that, due to radiative effects,
 $C$ {\it cannot} be significantly smaller than
$(\frac{1}{16\pi^2})^2$ in the hidden sector models.
In visible sector models, things are a bit more
complicated. Typical visible sector models include a so-called
messenger sector which contains a vector-like quark and lepton
superfields \cite{nelson}. By the aid of  messegner $U(1)$ gauge interaction,
the SUSY breaking dynamics gives rise to a mass splitting $\delta M$
in the messenger sector.
Subsequent radiative effects of the standard model gauge
interactions then lead to the
mass splitting $m_{\rm soft}\simeq \frac{\alpha}{4\pi}\delta M$
in the supersymetric standard model sector.
Since $\delta M$ is independent of the QCD coupling at
leading order, loops of the messenger sector particles will induce 
a QCD coupling-dependent vacuum energy density again at two loop order.
The corresponding dilaton potential energy is of order 
$(\frac{1}{16\pi^2})^2(\delta M)^2 M_m^2$ which is
of order $m_{\rm soft}^2 M_m^2$.
Thus in visible sector models, again due to radiative effects,
$C$ {\it cannot} be significantly smaller than order unity.

With the dilaton potential energy given in Eq. (27) in the early universe,
the Hubble expansion rate is  given by
\be
H\simeq C^{1/2} m_{\rm soft} M_m M_{\rm Pl}^{-1},
\ee
where $M_{\rm Pl}=2.44\times 10^{18}$ GeV.
Let us now examine whether the two conditions of Eq. (2)
for the relaxation of the axion misalignment  can be satisfied.
We consider first the case that $\Lambda_{QCD}<
m_{\rm soft}$ without any light quark.
 One can easily confirm that, as long as the complex parameters
contributing to $\theta_{\rm eff}$ are unchanged,
the axion VEV determined by the early axion potential
Eq. (4), i.e. $\theta_{\rm in}$, coincides with $\theta_{\rm eff}$. 
In the previous section, we have noted that 
all quark masses would become larger than $\Lambda_{QCD}$
if both of the Higgs soft masses $m_1^2$ and $m_2^2$
of Eq. (6)  receives a significantly large   
{\it negative} contribution from the energy density
in the early universe. This would make the Higgs  VEVs slide
down toward the flat direction $|H_u|=|H_d|\gg m_{\rm soft}$.
(Note that a large positive contribution leads to vanishing Higgs VEVs.) 
However, as was noted in Ref. \cite{thomas},
such a negative contribution arises usually
through supergravity interactions of the form $\frac{1}{M_{\rm Pl}^2}
\int d^2\theta d^2\bar{\theta} \, \phi\phi^*\chi\chi^*$ 
where $\chi$ is an inflaton-like field yielding
an energy density of order $H^2M_{\rm Pl}^2$.
For the above operators significantly alter 
the Higgs soft mass,  one needs $H$ to be comparable
to $m_{\rm soft}$.  Obviously then, we obtain $m_a\ll H$ 
since  $\Lambda_{QCD}< m_{\rm soft}\ll f_a$ in this case.

Let us consider the next case that  $\Lambda_{QCD}$ is still  less
than $m_{\rm soft}$ but now there exists a light quark.
The axion potential in this case is given by Eq. (5).
Using this axion potential and Eq. (28), we then find
\be
\frac{m_a}{H}\leq
5 \,C^{- 1/2} \left(\frac{4\times 10^{12} }{f_a}\right)
\left(\frac{10^5}{M_m}\right)\left(\frac{m_q}{10^{-2}}\right)^{1/2}
\left(\frac{\Lambda_{QCD}}{m_{\rm soft}}\right)^{3/2}\left(\frac{
m_{\rm soft}}{10^2}\right)^{1/2},
\ee
where all  numbers in the brackets denote the energy scales in  GeV unit.
The messenger scale $M_m$ can be as low as $10^5$ GeV
in the visible sector models for SUSY breaking.
However, in other type of models including the popular hidden sector models,
$M_m$ is typically much bigger than $10^5$ GeV.
As was mentioned, complex parameters contributing to 
$\theta_{\rm eff}$ in the early universe 
are required to have the values which are the same as 
the present one in order for 
the early axion VEV ($\theta_{\rm in}$) to coincide with the present VEV
($\theta_{\rm eff}$).
Real soft scalar masses $m_i^2$ escape from this requirement.
They might be significantly
bigger than the present values of order $10^2$ GeV due
to contributions from the radiation or other forms of energy density.
Note that if the early universe does not carry  quantum numbers 
which are carried by  some complex parameters,
a nonzero  energy density in such an early universe
would enhance  $m_i^2$ without affecting  those  complex parameters.
However, if the enhanced soft  mass-squared of Higgs fields
are significantly bigger
than the unaffected  $B\mu\sim 10^2$ GeV, it  makes
the Higgs VEVs to vanish.
This leads to $m_q=0$ and thus dramatically suppresses
the axion potential.

Taking into account the points discussed above, 
it is easy to note that, for a given value
of $f_a$, the ratio  $m_a/H$  becomes maximal
for $\Lambda_{QCD}$ comparable to $m_{\rm soft}$
which is of order $10^2$ GeV.
Then $m_a$ would  be large enough to push the axion field
toward the minimum 
if $f_a$ is in the range:  $f_a\leq (5 C^{-1/2})\times
 (10^5/M_m)\times (4\times 10^{12})$ GeV. 
Clearly in the hidden sector models with $M_m=M_{\rm Pl}$,
this range of $f_a$ does not overlap with the interesting
range
$f_a\gg 4\times 10^{12}$ GeV.
In visible sector models, the messenger scale $M_m$
corresponds to the scale of dynamical SUSY breaking
and thus may be as low as  $10^5$ GeV.
However as we have discussed, $C$ is at least
of order unity in visible sector models
and thus the axion mass is {\it not} large enough to relax the
axion misalignment for $f_a\gg 4\times 10^{12}$ GeV.

So far, we have argued that raising  $\Lambda_{QCD}$ 
up to the order of $m_{\rm soft}$ or below (while keeping
the complex parameters that contribute to $\theta_{\rm eff}$
unchanged) is not useful for relaxing the axion
misalignment when $f_a\gg 4\times 10^{12}$ GeV.
One might expect that raising  $\Lambda_{QCD}$ further up to  far above
$m_{\rm soft}$ can leads to an axion mass $m_a\geq H$.
However, the axion potentials Eq. (21) and 
Eq. (24) computed in the previous
section indicate that the axion mass is highly suppressed
for $\Lambda_{QCD}\gg m_{\rm soft}$.
This is mainly because 
the Higgs VEVs and the condensates
of the quarks, squarks, and the gluinos do {\it not} become of order 
$\Lambda_{QCD}$, but they all {\it vanish}
when $\Lambda_{QCD}\gg m_{\rm soft}$.
As a result,  instanton amplitudes for the axion potential
are suppressed by 
the powers of small Yukawa couplings
and also of the  small mass parameters  $\mu$ and $m_{\rm soft}$.

Before discussing the size of the axion mass, let us 
briefly discuss the axion VEV.
As can be noticed, the axion VEV determined by  
the axion potentials Eq. (21) and Eq. (24) differ 
from the $\theta_{\rm eff}$ of Eq. (25) and Eq. (26)
even when all complex parameters contributing $\theta_{\rm eff}$
in the early universe have the same values as the present one.
In such a case,  the misalignment angle 
$\delta\theta=\langle a/f_a\rangle-\theta_{\rm eff}$
for the MSSM is given by
\be
(\delta\theta)_{MSSM}=N_1 {\rm arg}(m_{1/2}A^*)+N_2 {\rm arg}(m_{1/2}B^*),
\ee
where $N_1$ and $N_2$ are appropriate integers of order unity.
Similarly for the NSSM,
we have
\be
(\delta \theta)_{NSSM}=N {\rm arg}(m_{1/2}A^*)
\ee
for an integer $N$ of order unity.
Although can be nonzero, the above misalgnment angle
is severely constrained by the neutron electric dipole moment \cite{edm}
as
\be
\delta\theta\leq 10^{-2}\sim 10^{-3}.
\ee
This would be small enough to raise  the cosmological bound
on $f_a$ up to the grand unification scale or
the Planck scale.  However as we mentioned, the axion mass
becomes too small to relax down the axion misalignment
to a small value of order $\delta\theta$.

Again using the axion potentials Eq. (21) and Eq. (24),
we find
$$
(\frac{m_a}{H})_{MSSM}\simeq 10^{-11} C^{-1/2} 
\left(\frac{4\times 10^{12}}{f_a}\right)
\left(\frac{10^5}{M_m}\right)\left(\frac{\mu}{10^2}\right)^{3/2}
\left(\frac{10^2}{\Lambda_{QCD}}\right)^{1/2},
$$
\be
(\frac{m_a}{H})_{NSSM}\simeq 10^{-5} C^{-1/2} (\lambda_1^3\lambda_2)^{1/2}
\left(\frac{4\times 10^{12}}{f_a}\right)\left(
\frac{10^5}{M_m}\right)
\left(\frac{A}{10^2}\right)^{1/2}
\left(\frac{\Lambda_{QCD}}{M_{\rm Pl}}\right)^{1/2}, 
\ee
where again the numbers in the brakets denote energy scales
in GeV unit.
The above results  obviously indicate that $m_a\ll H$.
Although only two models are explicitly considered for the case
$\Lambda_{QCD}\gg m_{\rm soft}$,
the huge suppression of the axion potential
seems to be quite  generic. In particular, adding more colored
particles leads to a further suppression, and thus not
helpful at all. 
We thus conclude that
the axion misalignment {\it cannot} be relaxed down to
a small value even when $\Lambda_{QCD}\gg m_{\rm soft}$.
In summary, the analysis made in this section indicates
that
a stronger QCD in the early universe is not 
useful for relaxing 
the axion misalignment.

\section{conclusion}

In this paper, we have examined whether the axion misalignment
can be relaxed down to a small value  by a stronger QCD in the early universe.
This would allow the axion scale to be of order the grand unification scale
or the Planck scale without any cosmological difficulty.
We discussed in somewhat detail the axion potentials
in the early universe, in particular
for the case that $\Lambda_{QCD}\gg m_{\rm soft}$.
Taking into account the dilaton potential energy
associated with a stronger QCD, our analysis
indicates that the two conditions  $m_a\geq H$ and $\langle a/f_a\rangle=
\theta_{\rm eff}$ (up to a small misalignment of order
$10^{-3}\sim 10^{-2}$) {\it cannot} be satisfied simultaneously. 
We thus conclude that a stronger QCD in the early universe 
is not  useful for raising the cosmological upper bound
of the axion scale.

\acknowledgements
This work is supported in part by KOSEF through CTP of Seoul National 
University (JEK,KC), KAIST Basic Science Research Program (KC),
KOSEF Grant 951-0207-002-2 (KC), Basic Science Research Institute 
Programs of Ministry of Education BSRI-96-2434 (KC) and BSRI-94-2418 
(JEK), the SNU-Nagoya Collaboration Program of Korea Research Foundation 
(JEK,KC), and the Ministerio de Educaci\'on y Ciencia of Spain under 
research grant (HBK).

\begin{table}[ht]
\centering
\caption{Quantum numbers of superfields and spurions in MSSM}
\vspace{7mm}
\begin{tabular}{|c|c|c|c|}
 & $U(1)_A$ & $U(1)_X$ & $U(1)_R$ \\
\hline\hline
$Q$          & 1    & 0    & 1    \\ \hline
$u^c$, $d^c$ & 1    & $-1$ & 1    \\ \hline
$H_u$, $H_d$ & 0    & 1    & 0    \\ \hline
$e^{-Y}$     & 12   & $-6$ & 6    \\ \hline
$Z_u$, $Z_d$ & $-2$ & 0    & 0    \\ \hline
$Z_{\mu}$    & 0    & $-2$ & 2    \\ \hline
$d^2\theta$  & 0    & 0    & $-2$ 
\end{tabular}
\label{table:1}
\end{table}

\begin{table}[ht]
\centering
\caption{Quantum numbers of superfields and spurions in NSSM}
\vspace{7mm}
\begin{tabular}{|c|c|c|c|c|}
 & $U(1)_A$ & $U(1)_X$ & $U(1)_{X^{\prime}}$ & $U(1)_R$ \\
\hline\hline
$Q$          & 1    & 0    & 0    & 1    \\ \hline
$u^c$, $d^c$ & 1    & $-1$ & 0    & 1    \\ \hline
$H_u$, $H_d$ & 0    & 1    & 0    & 0    \\ \hline
$S$          & 0    & 0    & 1    & 0    \\ \hline
$e^{-Y}$     & 12   & $-6$ & 0    & 6    \\ \hline
$Z_u$, $H_d$ & $-2$ & 0    & 0    & 0    \\ \hline
$Z_1$        & 0    & $-2$ & $-1$ & 2    \\ \hline
$Z_2$        & 0    & 0    & $-3$ & 2    \\ \hline
$d^2\theta$  & 0    & 0    & 0    & $-2$ 
\end{tabular}
\label{table:2}
\end{table}


\begin{figure}
\caption{Instanton graph for the axion potential Eq. (21) of the MSSM.
The solid lines with and without waves around 
the instanton denote the gluino and quark modes, respectively, 
while the dotted lines are the Higgs and squarks fluctuations.
The dark blobs represent the insertions of complex couplings
which are explicitly written in the graph.
The vertices not marked with couplings are the 
QCD gauge couplings.}
\end{figure}

\begin{figure}
\caption{Instanton graph for the axion potential Eq. (24)
of the NSSM. Again the solid lines are for fermion modes , the dotted
lines for boson fluctuations, and the dark blobs  for 
the inserted complex couplings.}
\end{figure}

\end{document}